\documentclass[%
 twocolumn,
 amsmath,amssymb,
 aps
]{revtex4-1}
\usepackage{xcolor}
\usepackage{comment}
\usepackage{graphicx}
\usepackage{tcolorbox}
\usepackage{upgreek} 
\usepackage{dcolumn}
\usepackage{bm}

\newcommand{\C}{{\mathcal{D}}}

\begin{document}

\preprint{APS/123-QED}

\title{Integer defects, flow localization, and bistability on curved active surfaces}
\author{Rushikesh Shinde$^{1}$, Rapha\"el Voituriez$^{2,3}$, Andrew Callan-Jones$^{1,*}$}
\affiliation{%
$^1$ Universit\'e Paris Cit\'e and CNRS, Laboratoire Mati\`ere et Syst\`emes Complexes, F-75013 Paris, France}
\affiliation{%
	$^2$ Sorbonne Universit\'e and CNRS, Laboratoire Jean Perrin, F-75005 Paris, France}
\affiliation{%
	$^3$ Sorbonne Universit\'e and CNRS, Laboratoire de Physique Th\'eorique de la Mati\`ere Condens\'ee, F-75005 Paris, France}
\email{andrew.callan-jones@u-paris.fr}


\date{\today}

\begin{abstract}
Biological surfaces, such as developing epithelial tissues,
exhibit in-plane polar or nematic order and can 
be strongly curved.  
Recently, integer (+1) topological defects have been identified as morphogenetic hotspots in living systems.  Yet, while +1 defects in active matter on flat surfaces are well-understood, the general principles governing curved active defects remain unknown. 
Here, we study the dynamics of integer defects in an extensile or contractile polar fluid on two types of morphogenetically-relevant substrates : (1) a cylinder terminated by a spherical cap, and (2) a bump on an otherwise flat surface. 
Because the Frank elastic energy on a curved surface generically induces a coupling to \emph{deviatoric} curvature, $\mathcal{D}$ (difference between squared principal curvatures), a +1 defect is induced on both surface types.  We find that $\mathcal{D}$ leads to surprising effects including localization of orientation gradients and active flows, and particularly for contractility, to hysteresis and bistability between quiescent and flowing defect states.
%
\end{abstract}

\maketitle

 Axi-symmetric (or very close to), active curved surfaces abound in living systems: the actin-myosin cortex underlying the cell membrane~\cite{SALBREUX2012536,Kelkar:2020uf}, and multi-cellular, epithelial structures such as tissue invaginations within the embryo and developing secretory glands are a few examples~\cite{Andrew:2010uq}. Orientational order, either polar or nematic, and active behavior is also found at the sub-cellular and tissue scales, and topological defects occur and have biological relevance \cite{Ardaseva:2022wa,Balasubramaniam:2022ty,PhysRevX.12.010501,Shankar:2022ta}.
 
For example, integer defects have been identified as morphogenetic organizing centers~\cite{Maroudas-Sacks:2021vg,Balasubramaniam:2022ty}, and are associated with outgrowths such as tentacles and the head in \emph{Hydra}~\cite{Maroudas-Sacks:2021vg,Ravichandran2024.04.07.588499}.
 It has also been found \emph{in vitro} that a +1 defect within a flat, confined cell or bacterial monolayer leads to out-of-plane dynamics~\cite{Guillamat:2022wy,Meacock:2021vq, Basaran:2022wg}.  These studies suggest a close connection between active +1 defects and non-flat geometries, yet little is known about the underlying physical principles~\cite{Metselaar:2019vl,Hoffmann:2022up}.  Given the deep connection between curvature and topological defects in equilibrium ordered surfaces~\cite{Bowick_Giomi_2009,TurnerVitelliNelson}, it is reasonable to expect that the active case presents an even deeper field of study. 
 
 Integer defects on flat, two-dimensional active fluids are known to be susceptible to flow-order instability. In vitro experiments in which a +1 defect in a myoblast monolayer was imprinted by boundary conditions (BCs) within a  circular region revealed transitions between non-flowing, aster and flowing spiral defects~\cite{Guillamat:2022wy}, confirming a theoretical prediction~\cite{Kruseetal}.  Spiral flows were found to be accompanied by mound formation.  A simulation study has been carried out to model this type of behavior~\cite{Hoffmann:2022up}, but analytical approaches to shed light on the fundamental physics are so far lacking. 
 	
 Coinciding with recent advances in imposing substrate curvature on cultured cell monolayers~\cite{Callens:2020wu,Schamberger:2023vi, luciano2024multiscale}, a few studies have modeled how fixed, but tunable curvature impacts spontaneous flows in active nematic surfaces~\cite{Bell:2022wx, Glentis:2022aa, Vafa:2022tw}.  This approach is amenable to in-depth analytical studies, yet to date how substrate geometry influences +1 defect dynamics has not been systematically investigated~\cite{Vafa:2022tw}.
 
 In this letter, we consider an achiral active polar fluid on two prototypical, biologically-motivated
axi-symmetric surfaces~\cite{Andrew:2010uq}: (1) a cylinder terminated by a spherical cap, and (2) a bump on an otherwise flat surface. 
The nematic elastic energy on curved surfaces includes a coupling to \emph{deviatoric} curvature, $\mathcal{D}$~\cite{Helfrich:1988vk}, defined here as the difference between squared azimuthal and meridional curvatures.
In contrast to flat surfaces where BCs may impose a net +1 topological charge, here we find that on axi-symmetric surfaces $\mathcal{D}$ can serve a similar function.


Our results can be summarized as follows. On a type (1) surface we find an aster defect is stable with respect to contractility and unstable to extensility, like on a flat surface~\cite{Kruseetal}, but unlike there the activity threshold is independent of system size and the resulting spiral-like order and flows are localized to regions of small $\mathcal{D}$. On a type (2) surface, because of the region where $\mathcal{D}<0$ (referred to here as the skirt), an aster is unstable at zero activity towards a spiral state and active flows are threshold-less, occurring for all non vanishing activity .  While extensility leads to spatially extended flows, contractility localizes flows, and if high enough extinguishes them and restores the aster. 
Surprisingly, with increasing curvature of the skirt, the spiral-to-aster transition shifts from continuous to discontinuous and contractility-driven localised flows exhibit hysteresis.

\begin{figure}[h!]
	\centering
	\includegraphics[width=0.42\textwidth]{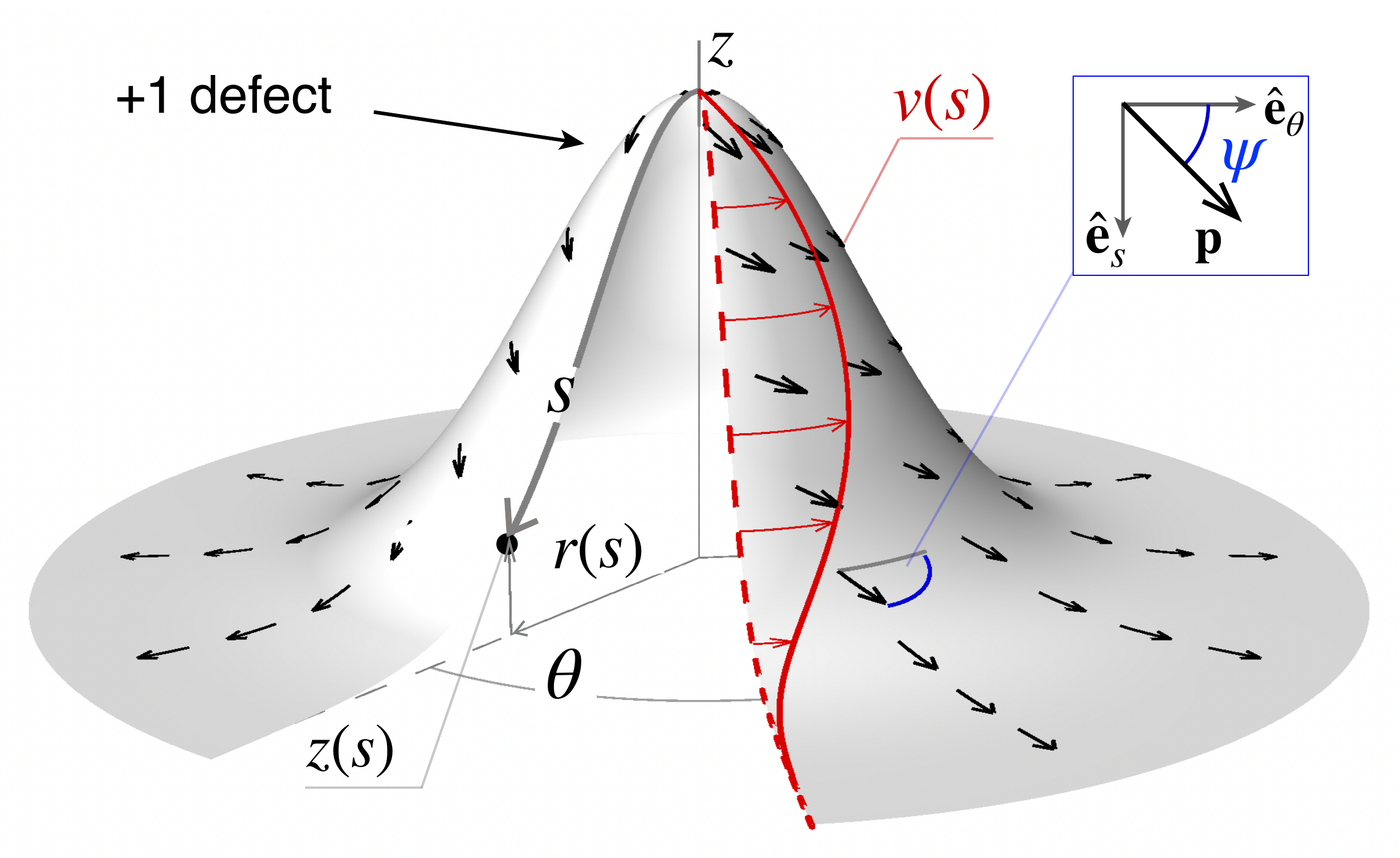}
	\caption{{\bf Illustration of a +1 defect in an active polar fluid on a surface of revolution}. The surface is defined by the coordinates $r(s)$ and $z(s)$. The polar field, 
	is indicated by black arrows.  Orthoradial flows $v(s)$ are represented by the red arrows and the red envelope curve.  }
	\label{fig:Fig1}
\end{figure}

We expect general biological relevance of this study.
On epithelial bud and tubular structures closed at one end, cells or actomyosin fibers are found to be aligned or tilted either axially \cite{Venzac:2018tj,Sanchez-Corrales:2018ub,Perez-Gonzalez:2021ta, Fernandez:2021ta} or 
ortho-axially \cite{Karner:2009tz}.
%
Such alignment necessarily encloses a +1 topological charge, making the physics predicted here  available to these  structures. \\ 
\noindent \emph{Model:} 
We consider an active polar fluid layer on a rigid, curved, axi-symmetric surface parametrized by the coordinates $s^{i}$, with $s^1=s$ the meridional arc length coordinate $s^2=\theta$ the azimuthal angle, as shown in Fig.~\ref{fig:Fig1}.  
The local order is described by a tangent vector $\mathbf{p}$ of unit length, and written as 
\begin{equation}
\mathbf{p}=\sin{\psi}\,\hat{\mathbf{e}}_s+\cos{\psi}\,\hat{\mathbf{e}}_\theta\,.
\label{eq:PdefPsi}
\end{equation}
Here, the unit tangent vectors $\hat{\mathbf{e}}_s$ and $\hat{\mathbf{e}}_\theta$ define the local orthonormal basis and
 $\psi(s,t)$ is the angle that $\mathbf{p}$ makes with $\hat{\mathbf{e}}_\theta$; see  Fig.~\ref{fig:Fig1} and SM, Sec.~1.  
Our goal here is to derive an equation for the dynamics of $\psi$ arising from elastic and active, as well as other out-of-equilibrium terms.  To do so, we first outline the geometry of the problem. 

%

Next, elastic effects are derived from a generalization of the two-dimensional Frank nematic free energy to include curvature.  For simplicity, we only keep terms that are up-down symmetric and achiral and include two terms: the first, with modulus $K$, couples to intrinsic (Gaussian) curvature and consists in replacing partial derivatives with covariant ones~\cite{NelsonPeliti_1987,LubenskyProst}; the second, with modulus $K_{\rm ex}$, couples to extrinsic curvature.  Thus, the energy is given by

%
\begin{align}
	F &=  \int \left(\frac{K}{2} \nabla_i p^j \nabla^i p_j +
	\frac{K_{\rm ex}}{2} \, p^j C_{ji} p_k C^{k i}\right) \sqrt{g}\,ds^1 ds^2 \nonumber \\
	&= \int \left(\frac{K}{2} \nabla_i p^j \nabla^i p_j+
	\frac{K_{\rm ex}}{4}\,\mathcal{D} \cos{(2\psi)}\right) \sqrt{g}\,ds^1 ds^2\,.
	\label{eq:FrankEnergy}
\end{align}
In the above, $\nabla_i p^j$ denotes the covariant derivative of the $j$ component of $\mathbf{p}$ with respect to the coordinate $s^i$, $i=1,2$;
$g=\mathrm{det}(g_{ij})$ is the determinant of the metric tensor; 
and $C_{ij}$ ($C^{ij}$) are the covariant (contravariant) components of the curvature tensor; see SI, Sec.~1 for details. We note that in the second line, dropping an unimportant constant term,
we have introduced the deviatoric curvature 
\begin{equation}
\C = (C_\theta{}^\theta)^2 - (C_s{}^s)^2\,.
\end{equation}
The extrinsic curvature term in Eq.~\ref{eq:FrankEnergy}, with two-fold symmetry about the normal, 
can be obtained from generic considerations of the bending energy of anisotropic membranes~\cite{Helfrich:1988vk}.  A related term with four-fold symmetry~\cite{Mbanga:2012tu} has, for simplicity, been neglected~\cite{FourFoldTerm}. 
We note that the two-fold term can also be obtained from a surface generalization of the Frank energy in the one constant approximation, which leads to $K_{\rm ex}=K$~\cite{Napoli_Extrinsic_curvature}. 

In the following we assume that $K_{\rm ex}>0$~\cite{NoteonKex1}.
It then follows that on a type (1) surface, where $\mathcal{D}$ tends to a positive constant far from the apex, a +1 aster ($\psi=\pi/2$) is energetically favored.  On a type (2) surface, the skirt region with $\mathcal{D}<0$ energetically favors a +1 vortex ($\psi=0$) defect. Thus, for an axi-symmetric (or very close to it) surface we expect a net +1 charge to be enclosed, whether the fluid is polar or nematic.
Based on these considerations, we assume that $\psi=\psi(s)$, meaning that a +1 defect is found at the surface apex $s=0$, and the problem is fully axi-symmetric. 

Activity and other, passive out-of-equilibrium effects in the fluid are described here by a simple constitutive law for the in-plane non-equilibrium tension tensor~\cite{PhysRevResearch.4.033158}:
\begin{equation}
\mathbf{t} = \zeta\Delta\mu\mathbf{p}\otimes\mathbf{p}+2 \eta \mathbf{u} +\frac{\nu}{2} \big( \mathbf{h} \otimes \mathbf{p} + \mathbf{p} \otimes \mathbf{h}  \big) + \frac{1}{2}\big( \mathbf{h} \otimes \mathbf{p} - \mathbf{p} \otimes \mathbf{h}  \big)\,,
\label{eq:TensionTensor}
\end{equation}
where $\otimes$ denotes the dyadic product.
Note that this tensor is defined such that $\mathbf{t}\cdot\hat{\boldsymbol{\nu}}\,dl$ is the internal force acting along a segment in the fluid of length $dl$ and perpendicular to a unit tangent vector $\hat{\boldsymbol{\nu}}$. 
In Eq.~\ref{eq:TensionTensor} $\zeta\Delta\mu$ is the coefficient of active tension, and positive in the contractile and negative in the extensile case; $\eta$ is the in-plane shear viscosity and $\mathbf{u}=u^{ij}\mathbf{e}_i\otimes\mathbf{e}_j$, $u^{ij}=\frac{1}{2}(\nabla^i v^j+\nabla^j v^i)$ is the strain rate associated with the velocity field $\mathbf{v}=v^i\mathbf{e}_i$ ; $\nu$ is the flow-alignment parameter~\cite{deGennesProst}, and $\mathbf{h}=-\delta F/\delta \mathbf{p}$ is the molecular field; and the final term is the antisymmetric tension, as required by rotational invariance~\cite{PhysRevResearch.4.033158}. 
These equations are identical to those for a three-dimensional active polar gel~\cite{Kruseetal,julicher2018hydrodynamic}, with the difference that $\zeta\Delta\mu$ and $\eta$ are surface quantities.  

The mathematical formulation is completed by specifying the dynamics of $\mathbf{p}$; for simplicity we only consider passive terms, and thus
\begin{equation}
\frac{D \mathbf{p}}{D t} = \frac{\mathbf{h}}{\gamma} - \nu \mathbf{u} \cdot \mathbf{p}\,,
\label{eq:polarity_dynamics}
\end{equation}
where
$D\mathbf{p}/Dt = \partial \mathbf{p}/\partial t + \mathbf{v} \cdot \nabla \mathbf{p} - \boldsymbol{\omega} \times \mathbf{p}$ is the co-moving, co-rotational derivative of $\mathbf{p}$, and   $\boldsymbol{\omega}=\frac{1}{2}\epsilon_{ij}\nabla^i v^j\,\mathbf{n}$ is the vorticity vector.  Finally,  $\gamma$ is the rotational viscosity. 
We focus on the case where the fluid consists of rod-like particles; passive nematic arguments~\cite{Helfrich_flow_alignment} suggest the alignment parameter $\nu<-1$, in agreement with measurements from different epithelial systems~\cite{Blanch-Mercader:2021vp,aigouy2010cell}. This will be important in what follows.

We note that in Eq.~\ref{eq:TensionTensor} we have not included a pressure-like term, as we only require off-diagonal components of $\mathbf{t}$.  A hypothesis on the compressibility of the layer is thus not required. Neglecting external forces tangent to the surface, such as friction or polar active self-propulsion~\cite{Glentis:2022aa}, and admitting only the possibility of circumferential flows, $\mathbf{v}=v^\theta \mathbf{e}_\theta$, the axi-symmetry of the problem implies that the total torque along $z$ acting on the sub-surface enclosed by a contour $r(s)=r$ is zero: 
	$\hat{\mathbf{z}}\cdot\oint_r \mathbf{X}\times (\mathbf{t}\cdot\hat{\mathbf{e}}_s)\,dl=0$.
Here, we have neglected any contribution from a non-equilibrium moment tensor, and have noted that the above equality holds identically for the equilibrium tensions and moments because of rotational invariance about $z$~\cite{PhysRevResearch.4.033158}.  We therefore find that the shear tension vanishes:
\begin{equation}
t^{s\theta}=0\,.
\label{eq:ZeroShear}
\end{equation}
This allows us to eliminate the strain rate $u^{s\theta}$ and therefore any explicit velocity dependence in Eq.~\ref{eq:polarity_dynamics}.  With some algebra, projecting Eq.~\ref{eq:polarity_dynamics} along $\mathbf{n}\times\mathbf{p}$ results in 
\begin{align}
\partial_t \psi &= \frac{1}{\tilde{\gamma}(\psi)}\left(K\nabla^2\psi
+\frac{1}{2} K_{\rm ex}\,\mathcal{D}\sin{(2\psi)}\right) \nonumber \\
&\hspace{1cm}+\frac{\zeta\Delta\mu}{4\tilde{\eta}(\psi)}(1+\nu\cos{(2\psi)})\sin{(2\psi)}\,,
\label{eq:PsiDynamics}
\end{align}
where $\tilde{\gamma}(\psi)>0$ and $\tilde{\eta}(\psi)>0$; see SM, Sec.~2 for details. 
%
Equation~\ref{eq:PsiDynamics} is the dynamical equation for $\psi$ we sought,
and it has two key, curvature-dependent differences compared with its flat space analogue~\cite{Kruseetal}: the Laplacian has been replaced with the Laplace-Beltrami operator, $\nabla^2\psi\equiv\frac{1}{\sqrt{g}}\partial_j(\sqrt{g}\,g^{ij}\partial_i\psi)$, encoding intrinsic curvature; and the presence of the coupling to extrinsic curvature, here related to $\mathcal{D}$.  These differences lead to surprising qualitative changes that we demonstrate below.

\noindent\emph{Linearized dynamics --- localization by curvature:}
It will prove useful to have the linearized version of Eq.~\ref{eq:PsiDynamics} about
aster, $\psi = \pi/2+\delta\psi$, and vortex, $\psi=\delta\psi$, defects. For small $\delta \psi$ we have
\begin{equation}
\frac{\partial \delta \psi}{\partial t} = \frac{K}{\gamma_{\pm}} \left( \nabla^2 \delta \psi \mp \frac{K_{\rm{ex}}}{K}\,\C \,\delta \psi
\right) +\frac{\zeta \Delta \mu (\nu \mp 1)}{2\eta} \delta \psi\,,
\label{eq:PsiDynamicsLinearized}
\end{equation}
where $\gamma_{\pm}^{-1}\equiv \gamma^{-1}+ (1\mp\nu)^2/4\eta$, and the upper (lower) sign here and above is for an aster (vortex). 
We remark on the passing resemblance of this equation to the Schr\"odinger wave equation~\cite{sakurai2020modern}, with the extrinsic curvature term akin to an external potential.  For constant $\mathcal{D}$, small deviations from aster order   vary on a length scale
\begin{equation}
l^{-2}=\frac{\zeta\Delta\mu \gamma_{+} ( 1-\nu)}{2 K\eta} + \frac{K_{\rm ext}}{K} \mathcal{D}\,.
\label{eq:CharacteristicLength}
\end{equation}
If $l^{-2}$ is positive the above equation implies exponential decay of $\delta\psi$.  In the linear regime flows induced by $\delta\psi$ will be seen to decay similarly. 

\noindent \emph{Localized flows on a type (1) surface:} 
The physics of an aster or vortex  defect in a 2D, initially quiescent, active fluid of rod-shaped constituents confined to a disk of diameter $D$ can be summarized as follows~\cite{Kruseetal}: a supercritical bifurcation occurs at a critical extensile activity $\zeta\Delta\mu_c\propto -D^{-2}$, beyond which the defect destabilizes towards a spiral state with active flows that span the disk.  This picture is significantly changed when the substrate is curved. 

\begin{figure}[h!]
\centering
\includegraphics[width=0.49\textwidth]{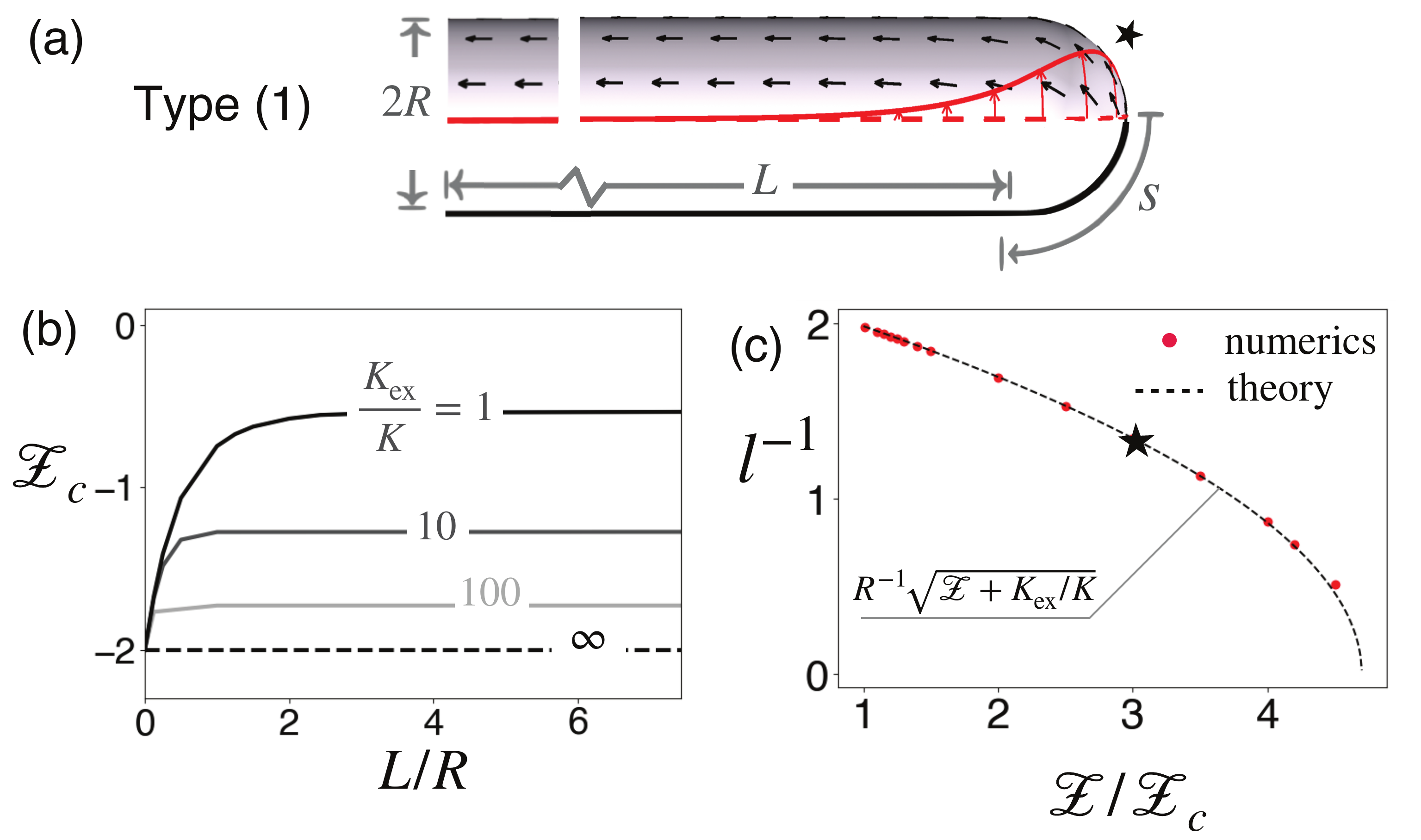}
\caption{ 
\textbf{Spontaneous localized flow for extensile activity.}
(a) Active polar fluid on a sphero-cylinder.  Above 
$\zeta\Delta\mu_c$, $\delta\psi(s)$ and $v(s)$ are non-zero and decay exponentially. 
(b) Dimensionless activity threshold $\mathcal{Z}_c\equiv\zeta\Delta\mu_c(1-\nu)\gamma_+ R^2/2\eta K$ versus $L/R$. 
(c)  Decay length $l^{-1}$ versus $\mathcal{Z}\equiv\zeta\Delta\mu(1-\nu)\gamma_+ R^2/2\eta K$.  Red points/Dashed line: numerical/analytical determination of $l^{-1}$.  The numerical value is found from the slope of $\log{v}$ vs $s$ (SI).  
 The polarity and flow profiles in (a) were obtained for the value of $\mathcal{Z}$ indicated by the star ($\star$). 
Parameters : $\nu=-2, \gamma=\eta=1, K_{\rm{ext}}/K=5, R=1, L=10$.  
 }
	\label{fig:Fig2}
\end{figure}

We first consider the fluid to be bound to a sphero-cylinder of radius $R$, as illustrated in Fig.~\ref{fig:Fig2}a. Far away from the cap and at a distance $L$ along the cylinder the orientation is assumed fixed to be $\psi=\pi/2$~\cite{NeumannCondition}.  By symmetry the initial configuration is therefore aster-like.  We look for the critical activity for an aster-to-spiral (A-S) transition. We show that above this value spiral order and the attendant flows are localized to the cap; a second transition occurs at higher extensile activity to the familiar bend-dominated distortion along the length of the cylinder (not shown; see Refs.~\cite{NapoliTurzi,Al-Izzi_Morris_2023}).  To find $\zeta\Delta\mu_c$, we consider the linearized equation, Eq.~\ref{eq:PsiDynamicsLinearized}, with the upper signs. The deviatoric curvature is everywhere non-negative:
on the hemispherical cap $\mathcal{D}=0$ and on the cylindrical surface $\mathcal{D}=1/R^2={\rm const}$.  It follows that the aster is linearly stable with respect to contractility, as on a flat surface~\cite{Kruseetal} --- using the quantum mechanical analogy, there are no ``negative energy states''.  Thus, aster instability only occurs for extensile activity, which is the case of interest for the rest of this section.  A non-trivial stationary solution to Eq.~\ref{eq:PsiDynamicsLinearized} in analytical form can then be obtained:
%
\begin{equation}
	\delta \psi = 
\begin{cases}
P_n[\cos{(s/R)}]\,,& 0 < s \le R\,; \\[.2cm]
A e^{-s/l_c}+B e^{s/l_c}\,, & R < s\le L\,.	
\end{cases}
\end{equation}
Here, $A$ and $B$ are unknown constants; $P_n$ is the Legendre polynomial of order
$n = \big(\sqrt{1-4 \mathcal{Z}_c}-1\big)/2$, $n\in\mathbb{R}$; and
$\mathcal{Z}_c\equiv\zeta\Delta\mu_c(1-\nu)\gamma_+ R^2/2\eta K = -n(n+1)$ is the dimensionless activity threshold. 
The length $l_c$, as defined in Eq.~\ref{eq:CharacteristicLength} and at threshold, is in the current geometry given by
$ l_c^{-2}  = R^{-2}\,(\mathcal{Z}_c+K_{\rm ex}/K)$.  Applying the boundary condition $\delta\psi(s=L+\pi R/2)=0$ and requiring continuity of $\delta\psi$ and $\partial_s\delta\psi$ at $s=\pi R/2$ allow determination of $A$, $B$, and $\mathcal{Z}_c$ 
; see Fig.~\ref{fig:Fig2}b.  For large $L/l_c$, $B\simeq 0$, and $\mathcal{Z}_c$ is obtained by solving
\begin{equation}
(n+1) P_{n+1}(0)+\sqrt{K_{\rm ex}/K-n(n+1)}\,P_n(0)=0\,.
\label{eq:EigenvalueEquation}
\end{equation}
%
The first root, $n_{c}$, corresponding to the activity threshold, is less than 1, i.e., $\mathcal{Z}_c>-2$. For $K_{\rm ex}=K$, $\mathcal{Z}_c\approx -0.53$; 
with increasing $K_{\rm ex}/K$, $\mathcal{Z}_c\to-2$ (Fig.~\ref{fig:Fig2}b). In this limit, the cylinder curvature completely screens $\delta\psi$, and spiral reordering is confined to the cap.  

Above threshold the flow $v=\hat{\mathbf{e}}_\theta\cdot\mathbf{v}$ is non-zero and is obtained by integration of Eq.~\ref{eq:ZeroShear}; see SM, Sec.~3.  It vanishes at $s=0$, is maximal along the hemispherical cap and decays exponentially along the cylinder with characteristic length, $l$;~see SM, Sec.~4.  This exponential decay holds well-above threshold; see Fig.~\ref{fig:Fig2}c.  
The steady state that results 
is akin to a bound state of a quantum particle with spatial extent $l$ that depends on $R$ but not on the system size $L$.

\begin{figure}[h!]
\includegraphics[width=0.49\textwidth]{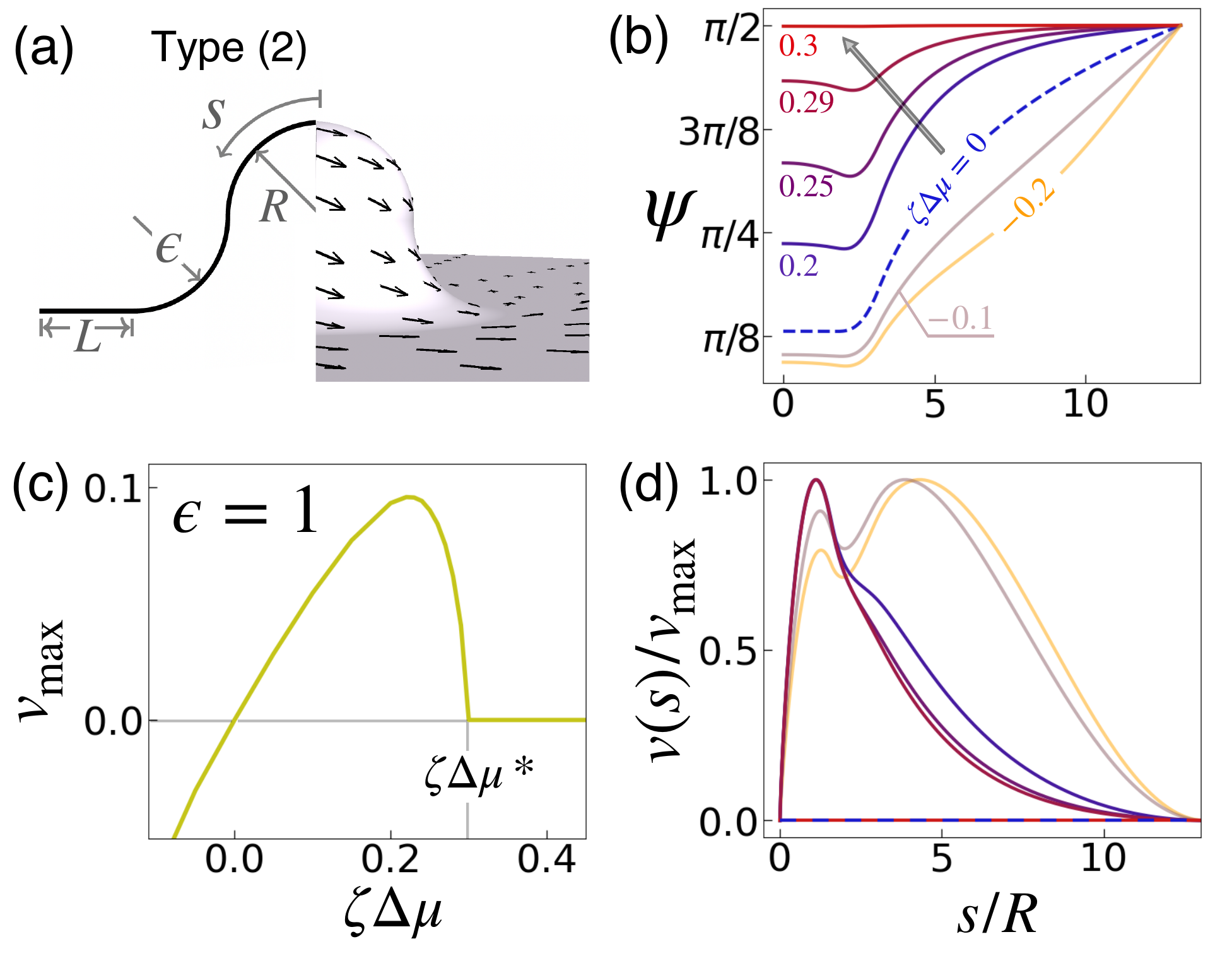}
\caption{ \textbf{Spherical bump with low skirt curvature.} (a) Illustration of surface, with non-aster order in the zero activity state.
(b) $\psi(s)$ profiles for different activity values. Contractility restores aster order, and the S-A transition is continuous. Extensility further destabilizes aster order. (c)  The maximal value of $v(s)$ versus activity.  It is defined as $v_{\rm max}=\mathrm{sgn}(\zeta\Delta\mu)\,\mathrm{max}[v(s)]$.  Flows are threshold-less and smoothly vanish at a value $\zeta\Delta\mu^*$.
(d) Contractile flows are localized whereas extensile flows span the system size. Parameters : $\nu=-2, \gamma=\eta=1, K_{\rm{ext}}/K=1, R=\epsilon=1, L=10$.  }
\label{fig:Fig3}
\end{figure}

\noindent \textit{Localized flows on a type (2) surface:} 
As a second illustration of the influence of extrinsic curvature on active +1 defects, 
%
we consider a hemispherical bump of radius $R$ that is smoothly joined to a surrounding flat surface via a skirt (part of a torus) of major radius $R+\epsilon$ and minor radius $\epsilon$; see Fig.~\ref{fig:Fig3}a. At any point on the skirt it can be readily shown that $\mathcal{D}\le 0$.
%
For simplicity, we assume that aster conditions, $\psi=\pi/2$, are imposed at the edge of the domain.    This gives rise to frustrated order and bistability, which are new results in this work, though other choices of BCs can be studied~\cite{SphericalBumpTangentialOrder}.  Bistability is specific to frustration and would not occur for all choices.
At zero activity, $\mathcal{D}$ will drive the instability of an aster towards spiral-like order through a pitchfork bifurcation, if the skirt is sufficiently curved; see SM, Sec.~5.  In the following, we assume that the zero activity state is spiral-like, with a non-vanishing gradient in $\psi$; see Fig.~\ref{fig:Fig3}b.  

Turning on activity, the order profiles are altered: 
contractility tends to restore the aster state, and extensility expands the spiral state (Fig.~\ref{fig:Fig3}b), consistent with their respective stabilizing and destabilizing effects on bend distortions~\cite{edwards2009spontaneous}.  Because of the director gradients in the passive state, the active flows are threshold-less~\cite{green2017geometry}, increase monotonically with extensility and non-monotonically with contractility, and vanishing at some value $\zeta\Delta\mu^*$, at which a spiral-to-aster (S-A) transition occurs (Figs.~\ref{fig:Fig3}b, c). 
Furthermore, flows become more localized in the skirt with increasing contractility: away from the bump $v(s)$ decays exponentially with characteristic length given by Eq.~\ref{eq:CharacteristicLength} with $\mathcal{D}=0$; see Figs.~\ref{fig:Fig3}d and SM, Sec.~6.
Overall, the possibility of contractility-generated active flows near integer defects, previously not found in flow-aligning ($\nu<-1$) active fluids on flat surfaces~\cite{Kruseetal}, is due to the spiral order favored by $\mathcal{D}<0$,
and is traced back to the elastic energy dependence on extrinsic curvature. 

\noindent \textit{First order transition at high} $|\C|$ : 
For greater inner curvature $1/\epsilon$ we were surprised to find that the S-A and A-S transitions become discontinuous. 
Starting in the spiral state --- the yellow curve in Fig.~\ref{fig:Fig4}, and referred to as S1---, with increasing contractility numerical integration of Eqs.~\ref{eq:ZeroShear} and \ref{eq:PsiDynamics} shows that the active flows first increase and then suddenly vanish at a value $\zeta\Delta\mu_1$ (Fig.~\ref{fig:Fig4}a). For large $1/\epsilon$ the skirt favors vortex-like order, which is maintained on the bump (Fig.~\ref{fig:Fig4}b).  With increasing contractility, the angle $\psi$ becomes close to $\pi/2$ everywhere on the flat part of the surface, and close to zero on the bump.  This can be understood by noting from Eq.~\ref{eq:PsiDynamicsLinearized}, that for $\nu<-1$, contractility stabilizes both aster \emph{and} vortex order. 
At $\zeta\Delta\mu_1$ this geometry-induced frustration in the order is suddenly relieved, and the state S1 transitions to an aster.   
Hysteresis is also observed: as contractility is then decreased, the flows remain zero until a lower threshold $\zeta\Delta\mu_{2}$ at which aster order is continuously lost by a symmetry-breaking instability towards a different spiral state --- referred to as S2---, and the order becoming vortex-like in the skirt (Fig.~\ref{fig:Fig4}c). The final step is a snap-like transition of S2 to S1 at $\zeta\Delta\mu_3$ (Figs.~\ref{fig:Fig4}a and c).

Our results for the bump surface are summarized in the phase diagram in Fig.~\ref{fig:Fig4}d. The key feature is the tri-critical point (T) that separates the regions where the A-S and S-A transitions are supercritical, from where they are subcritical and bistability can occur.  The bistable region, bounded by the curves I and III
is further divided into subregions: between I and II the aster and S1 states coexist; and between  II and III the two flowing states, S1 and S2, coexist.
We note that our findings presented in Fig.~\ref{fig:Fig4}d are not unique to our specific choice of bump surface in Fig.~\ref{fig:Fig3}a.  
We found similar behavior, notably the presence of a tri-critical point in the phase behavior, in other surfaces of revolution, namely the revolved bump function and a bulb-tipped cylinder, which combines features of surfaces of types (1) and (2); see SM, Sec.~7.  This suggests generic behavior shared by surfaces containing regions where $\mathcal{D}<0$.  

\noindent{\emph {Discussion:}
	We have studied +1 defects in active polar fluids on curved surfaces.  
We found that extrinsic curvature qualitatively changes the known physics of +1 active defects in flat 2D fluids: 
it provides spatial control of active flows and hysteresis between flowing and non-flowing states. 
Our work joins that of other
recent studies on bistability in active nematics~\cite{Bell:2022wx,lavi2024nonlinear}.   

We expect that the physics discovered here is biologically relevant. 
Rotational flow of cells in the bulb-shaped acinus of organoid-derived mammary epithelial tubes has been observed, with an amplitude decaying with distance along the tube~\cite{Fernandez:2021ta}. 
Our model might provide an explanation for this behavior. 

Here, we have focused on active fluids that are polar by virtue of the assumption of an axi-symmetric, +1 defect imposed by substrate geometry. For simplicity we have not included  terms in the equations of motion that break the $\mathbf{p}\to-\mathbf{p}$ symmetry.  Furthermore, we have also not included substrate effects such as an active polar traction force~\cite{Glentis:2022aa}, couplings that break up-down layer symmetry, and passive friction with the substrate~\cite{NoteonFriction}.  Following other recent work~\cite{Vafa_PRE_2024}, we have pursued a minimal model on a fixed surface to show how curvature augments the physics of active defects.  Future work will more fully examine the parameter space that encompasses epithelial features not considered here.

Finally, we expect our work to be relevant to active nematics, as well. 
On axi-symmetric surfaces the curvature deviator $\mathcal{D}$, by favoring radial or circumferential alignment according to the Frank energy, Eq.~\ref{eq:FrankEnergy}, results in an enclosed +1 charge.  It remains to be
seen whether in such situations a single +1, two +1/2, or some other combinations of defects is selected. 

\onecolumngrid
\begin{center}
	\begin{figure}[h]
		\includegraphics[width=.76\textwidth]{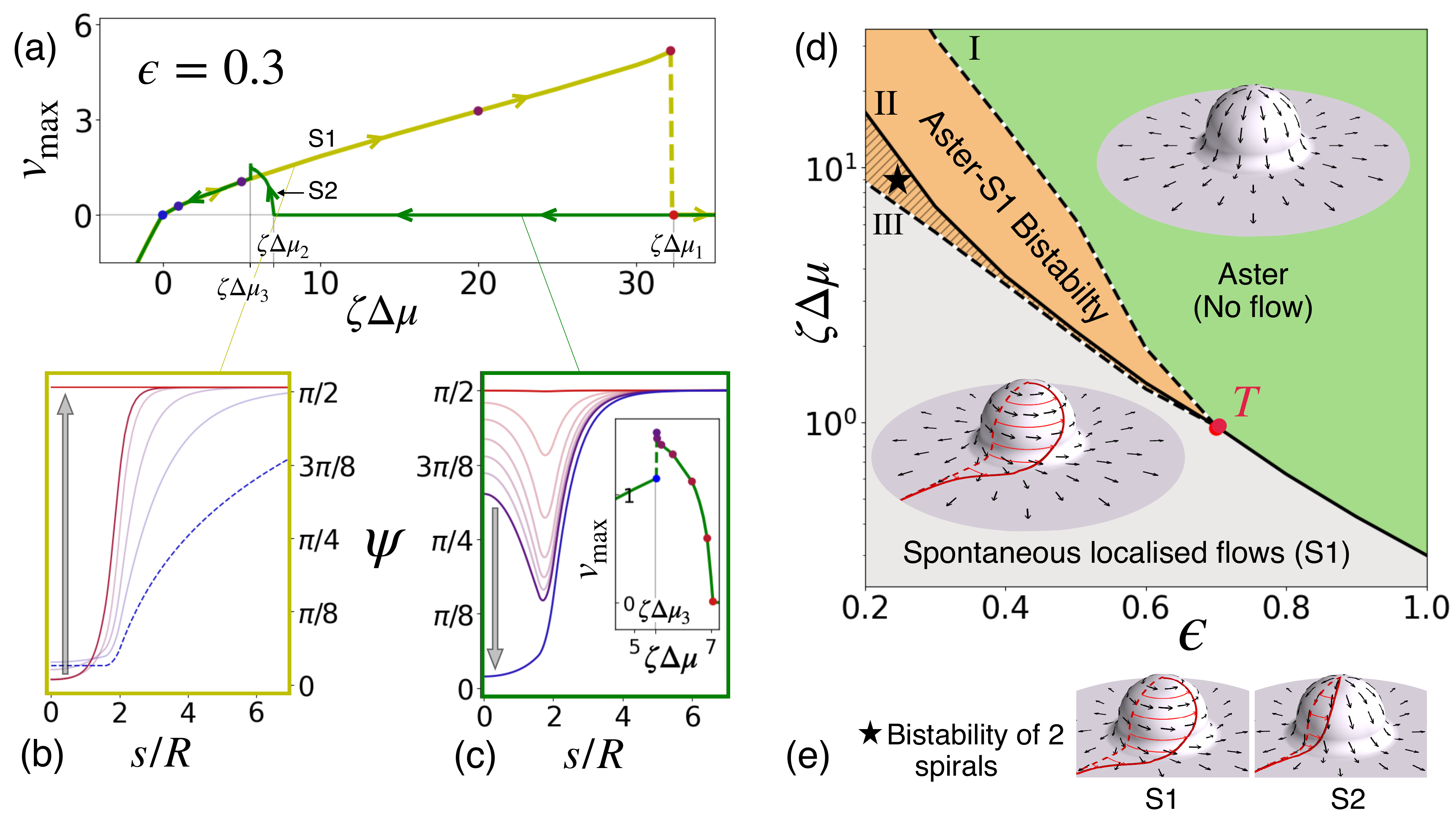}
		\caption{ \textbf{Spherical bump with high skirt curvature: bistability of spiral and aster states. }
		(a) The maximal value of $v(s)$ versus contractile activity.  Hysteresis and discontinuous transitions are found numerically, and are shown here.
		(b) The $\psi(s)$ profiles, color-coded according to the points in (a), show a discontinuous S-A transition with increasing $\zeta\Delta\mu$.
		(c) The A-S is also discontinuous, but is preceded by a continuously deformed aster state prior to snapping. The corresponding values of $v_{\rm max}$ along the the fin-shaped curve in (a) are given in the inset.
		(d)  Phase diagram in the $\zeta\Delta\mu$-$\epsilon$ plane.  A tri-critical point (T) is shown, and for skirt radii less than $\epsilon_{\rm T}$ the S-A and S-A transitions are first-order, indicating bistability of the spiral and aster.
		The dashed lines indicate discontinuous transitions, whereas the solid line continuous ones. (e)  Close-up of polarity and flow profiles for S1 and S2 spiral states.
			Parameter values:	$\nu=-2, \gamma=\eta=1, K_{\rm{ext}}/K=1, R=1, \epsilon=0.3, L=10$. 
		}
		\label{fig:Fig4}
	\end{figure}
\end{center}
\twocolumngrid

\begin{acknowledgments}
RS acknowledges PhD funding from Ecole Doctorale Ile de France (EDPIF, contract number 1960999223S01).
	RV is funded by ERC synergy grant SHAPINCELLFATE.
	We also thank Jacques Prost for his careful reading and insightful comments on the manuscript. 
\end{acknowledgments}
\bibliography{references_acj}

\begin{thebibliography}{52}
\bibitem [{Neu()}]{NeumannCondition}%
  \BibitemOpen
  \href@noop {} {}\bibinfo {note} {We note that for $L\gg R$ free boundary
  conditions on $\psi$ do not qualitatively change the results to
  follow.}\BibitemShut {Stop}%
\bibitem [{Sph()}]{SphericalBumpTangentialOrder}%
  \BibitemOpen
  \href@noop {} {}\bibinfo {note} {The case of $\psi=0$ conditions does not
  lead to new physics as compared with Ref.~\cite{Kruseetal}. If we impose free
  conditions, $\partial_s\psi=0$, we find that bistability between spiral and
  vortex states occurs over a range of contractilities, similar to that between
  spiral and aster states, when $\psi=\pi/2$ is fixed at the boundary. For
  simplicity we focus on these latter conditions.}\BibitemShut {Stop}%
\bibitem [{Not({\natexlab{b}})}]{NoteonFriction}%
  \BibitemOpen
  \href@noop {} {} ({\natexlab{b}}),\ \bibinfo {note} {friction, with
  coefficient $\xi$, combines with layer viscosity $\eta$ to yield the length
  $l_\eta = \sqrt{\eta/\xi}$. For phenomena occurring on lengths less than
  $l_\eta$, friction can be ignored.}\BibitemShut {Stop}%
\end{thebibliography}
\end{document}


\maketitle

\section{Some basic differential geometry}
\label{sec:DiffGeo}
We briefly define the geometrical quantities that are used in the main text.
First, we consider a surface endowed with metric $g_{ij}$ and curvature $C_{ij}$ defined as
\begin{align}
g_{ij}&=\mathbf{e}_i\cdot\mathbf{e}_j \\
C_{ij} &=\mathbf{e}_i\cdot \partial_j \mathbf{n}\,.
\end{align}
In these expressions, the tangent basis is
\begin{equation}
\mathbf{e}_i= \partial_i \mathbf{X}\,;
\end{equation}
partial derivatives are with respect to one of the two coordinates $s^i$, $i=1,2$ parametrizing the surface; and
$\mathbf{X}$ is the position vector of a point on the surface in the embedding space. 
Furthemore, $\mathbf{n}$ is the unit normal vector to the surface, given by 
\begin{equation}
\mathbf{n}=\frac{1}{2}\epsilon^{ij}\mathbf{e}_i\times\mathbf{e}_j\,,
\end{equation}
which defines the Levi-Civita tensor $\epsilon^{ij}$.  
Unit tangent vectors are denoted 
\begin{equation}
\hat{\mathbf{e}}_i\equiv \mathbf{e}_i/|\mathbf{e}_i|\,. 
\end{equation}
  Applying Einstein summation convention, the dual 
  tangent vectors are $\mathbf{e}^i \equiv g^{ij} \mathbf{e}_{j}$, where $g^{ij}$ is the inverse of the metric. Indices of a tangent vector $\mathbf{a}= a^{i}\mathbf{e}_i$, or second-rank (enough for our purposes) tensor  $\mathbf{A}=A^{ij}\mathbf{e}_i\otimes\mathbf{e}_j$, can be raised or lowered  by contraction with $g^{ij}$ or $g_{ij}$.
  
 Finally, the covariant derivative of a vector $\mathbf{a}$ is
 defined as
 $
 \nabla_i a^j = \mathbf{e}^j\cdot \partial_i \mathbf{a}\,,
$
and can be expressed in terms of the Christoffel symbols, $\Gamma_{ij}^k = \mathbf{e}^k \cdot \partial_i \mathbf{e}_j$, as
\begin{equation}
\nabla_i a^j 
=\partial_i a^j +\Gamma^{j}_{ik} a^k\,.
\label{eq:VectorCovDeriv}
\end{equation}
Note that the above definition assumes that $\mathbf{a}$ is a vector that is tangent to the surface.
\section{Derivation of $\psi$ dynamics : Equation 7 in the main text}
We next provide details on the derivation of Equation 7 in the main text, which describes the dynamics of the orientation angle $\psi$ on an axi-symmetric surface.  \\

\noindent \textbf{Parameterization of a surface using in-plane co-ordinates :} 
An axi-symmetric surface can be parameterized by using in-plane polar co-ordinates: arc length $s$ and azimuthal angle $\theta$; see Fig.~1 of the main text. Using these parameters, the position vector of a point on this surface in the standard Cartesian basis can be defined as $\mathbf{X}=[r(s) \cos{\theta}, r(s)\sin{\theta}, z(s)]^T$. From the definition of arc length $s$, it follows that $z'(s)^2+r'(s)^2=1$, where primes denote differentiation with respect to $s$.  Using the definitions given in Sec.~\ref{sec:DiffGeo} the tangent vectors are
\begin{align}
\mathbf{e}_\theta &= [r'(s)\cos{\theta},r'(s)\sin{\theta},z'(s)]^T\,,  \\ 
\mathbf{e}_s &= [-r(s)\sin{\theta},r(s)\cos{\theta},0]^T \,.
\end{align}
The metric tensor $g_{ij}=\mathbf{e}_i\cdot\mathbf{e}_j$ and the Levi-Civita tensor $\epsilon_{ij} = \mathbf{n} \cdot (\mathbf{e}_i \times \mathbf{e}_j)$ are given by
\begin{equation}
g_{ij} = \begin{bmatrix}
1 & 0 \\
0 & r(s)^2\\
\end{bmatrix} \hspace{20pt}{\rm{and}} \hspace{20pt} \epsilon_{ij} = \sqrt{g} \begin{bmatrix}
0 & 1 \\
-1 & 0\\
\end{bmatrix}\,.
\end{equation}
Here, we have used $g$ as shorthand notation for the determinant of the $g_{ij}$ matrix. In a contravariant basis, the above tensors are written as
\begin{equation}
g^{ij} = \begin{bmatrix}
1 & 0 \\
0 & 1/r(s)^2\\
\end{bmatrix},\hspace{40pt} \epsilon^{ij} = \frac{1}{\sqrt{g}} \begin{bmatrix}
0 & 1 \\
-1 & 0\\
\end{bmatrix}.
\end{equation}
Next, the normal vector $\mathbf{n}$ is given by
\begin{align}
\mathbf{n} &=  [-z'(s)\cos{\theta},-z'(s)\sin{\theta},r'(s)]^T\,
\end{align}
and the curvature is given by
\begin{equation}
C_{ij}=\begin{bmatrix}
r'(s) \, z''(s) - z'(s)  \, r''(s)& 0 \\
0 &  r(s) z'(s) \\
\end{bmatrix}.
\end{equation} \\
Finally, in computing the covariant derivative of a tangent vector $\mathbf{a}$
the only non-zero Christoffel symbols are 
\begin{equation}
   \Gamma_{\theta s}^{\theta } = \frac{r'(s)}{r(s)}, \hspace{20pt} \Gamma^{s}_{\theta \theta} =  -r'(s)r(s). 
\label{eq:Christoffel_symbols}
\end{equation}

\noindent \textbf{Dynamics of a +1 active defect :}  We next consider the polarity dynamics associated with a +1 defect in an active polar fluid surface. The defect center is found along the axis of symmetry of the surface, i.e., at $s=0$. 
Assuming rotational symmetry, the polarity field can be written as
\begin{equation}
\mathbf{p}=\sin{\psi(s,t)}\,\hat{\mathbf{e}}_s+\cos{\psi(s,t)}\,\hat{\mathbf{e}}_\theta\,
\label{eq:PdefPsi}
\end{equation}
and the flow field, which is orthoradial and axi-symmetric, is
\begin{equation}
\mathbf{v}=v^\theta(s,t)\,\mathbf{e}_\theta\,.
\label{eq:OrthoradialFlow}
\end{equation}
The dynamics of the polarity vector is described by the following equation (Eq.~5 in the main text)  :
\begin{equation}
    \frac{D \mathbf{p}}{D t} = \frac{\mathbf{h}}{\gamma} - \nu (\mathbf{u} \cdot \mathbf{p}).
    \label{eq:dynamics_of_polarity}
\end{equation}
Here, the co-rotational Lagrangian derivative is $D\mathbf{p}/Dt = \partial \mathbf{p}/\partial t + \mathbf{v} \cdot \nabla \mathbf{p} - \boldsymbol{\omega} \times \mathbf{p}$. On the right hand side of the above equation, the first term includes the molecular field $\mathbf{h} = - {\delta F}/{\delta \mathbf{p}}$ and the rotational viscosity $\gamma$. The second term represents the coupling between the polarity dynamics and the velocity field, where $\nu$ is known as the flow-alignment parameter---which is dimensionless---, and $\mathbf{u}=\big( \nabla \mathbf{v} +  \nabla \mathbf{v}^{\rm{T}} \big)/2$ is the strain rate tensor. Due to the assumed symmetry of the flow profile, Eq.~\ref{eq:OrthoradialFlow}, the only non-zero component of $\mathbf{u}$ is the shear component, $u^{\theta s}$.  
With the aid of Eq.~\ref{eq:VectorCovDeriv}, this component can be simplified as follows:
\begin{align}
    u^{\theta s} &= \frac{1}{2} \left( \nabla^s v^\theta + \nabla^\theta v^s \right), \notag \\
    &= \frac{1}{2} \left( \partial_s v^\theta + \frac{r'}{r} v^\theta -\frac{r'r}{r^2} v^\theta \right), \notag\\
    &=\frac{1}{2} \partial_s v^\theta.
    \label{eq:ShearStrainITOvtheta}
\end{align}
Next, we project Eq.~\ref{eq:dynamics_of_polarity} 
along $\mathbf{p}$ and the perpendicular tangent direction $\mathbf{p} \times \mathbf{n}$. This gives
\begin{align}
    \mathbf{p} \cdot \frac{D \mathbf{p}}{D t} &= \frac{h_\parallel}{\gamma}- \nu \mathbf{p} \cdot (\mathbf{u} \cdot \mathbf{p} ), \label{eq:Streching_Dynamics} \\
    (\mathbf{p} \times \mathbf{n} ) \cdot \frac{D \mathbf{p}}{D t} &= \frac{h_\perp}{\gamma}- \nu (\mathbf{p} \times \mathbf{n}) \cdot (\mathbf{u} \cdot \mathbf{p} )\,.
 \label{eq:Rotational_Dynamics}
\end{align}
Here, we have defined $h_{\parallel}$ and $h_{\perp}$ via $\mathbf{h} = h_\perp (\mathbf{p} \times \mathbf{n})  + h_\parallel \mathbf{p}$. Since the fluid is assumed to be deep in the ordered state, we have $|\mathbf{p}|=1$. This implies $\mathbf{p} \cdot D\mathbf{p}/Dt = 0$. This identity can be used to simplify the equation \eqref{eq:Streching_Dynamics}, which gives 
\begin{equation}
     h_\parallel = \nu \gamma \sqrt{g} u^{\theta s} \textrm{sin}\big(2 \psi \big). \label{eq:strain_ustheta_relation}
\end{equation}
Equation \eqref{eq:Rotational_Dynamics} can then be simplified as follows. The left-hand side is explicitly given by 
\begin{align}
     (\mathbf{p} \times \mathbf{n} ) \cdot \frac{D \mathbf{p}}{D t} &= \frac{\partial \psi}{ \partial t} + (\mathbf{p}\times \mathbf{n}) \cdot \big(  \mathbf{v} \cdot \nabla \mathbf{p} \big) - (\mathbf{p}\times \mathbf{n} )\cdot (\boldsymbol{\omega} \times \mathbf{p} )\\
     &= \frac{\partial \psi}{ \partial t} + (\mathbf{p}\times \mathbf{n}) \cdot (v^\theta \nabla_\theta p^j \mathbf{e}_j ) + \boldsymbol{\omega} \cdot \mathbf{n} \,,
\end{align}
In order to simplify the second term on the right hand side above, we first write $\mathbf{p}\times \mathbf{n}$ as $ = - \sin{\psi} \hat{\mathbf{e}}_\theta + \cos{\psi} \hat{\mathbf{e}}_s$. Then, with the expression for $\mathbf{p}$, Eq.~\ref{eq:PdefPsi}, and some algebra, this 
term is shown to be equal to $-v^\theta r'$.  The third term on the right hand side of the above equation, with the definition of the vorticity $\omega_n = \boldsymbol{\omega}\cdot\mathbf{n}= \epsilon_{ij} \nabla^i v^j/2$ yields $\omega_n= v^\theta r'+ \sqrt{g} u^{s\theta} = (r^2 v^\theta)'/2r$. These considerations lead to
\begin{equation}
    (\mathbf{p} \times \mathbf{n} ) \cdot \frac{D \mathbf{p}}{D t} = \frac{\partial \psi }{\partial t} +\frac{\sqrt{g}}{2}\partial_s v^\theta
\label{eq:co-rotational_derivative_projected}
\end{equation}
Having thus addressed the left hand side of Eq.~\ref{eq:Rotational_Dynamics}, we next tackle the right hand side. 
\noindent A straightforward calculation gives $(\mathbf{u} \cdot\mathbf{p}) \cdot (\mathbf{n} \times \mathbf{p} ) = \cos{2 \psi} u^{\theta s} \sqrt{g}/2$. Substituting this expression into Eq.~\ref{eq:Rotational_Dynamics} and using Eq.~\ref{eq:co-rotational_derivative_projected}, we obtain
\begin{align}
    \frac{\partial \psi }{\partial t}  = \frac{h_\perp}{\gamma} - \sqrt{g} \, \Big( 1+ \nu \, \textrm{cos} (2 \psi) \Big) u^{s \theta}\,.
 \label{eq:simplified_rotational_dynamics}
\end{align}
From this equation, we can obtain an autonomous dynamical equation for the angle $\psi$, provided we can express the 
 shear rate $u^{s\theta}$ as a function of $\psi$.  
 To do so,  we neglect the friction between the liquid film and the substrate, which, given the axi-symmetry of the problem, implies that the shear deviatoric tension $t^{s\theta}$ vanishes (see main text). By noting that this quantity is given by

\begin{equation}
  t^{s\theta}=  2 \eta u^{s \theta} + \zeta \Delta \mu \, \, p^\theta p^s + \frac{\nu}{2} \Big( p^\theta h^s + p^s h^\theta \Big) +\frac{1}{2}  \big( h^s p^\theta - h^\theta p^s \big)\,,
\end{equation}
(see main text, Eq.~4), and writing the 
molecular field 
as
\begin{equation}
\mathbf{h} = \big( h_\parallel \cos{\psi} - h_{\perp} \sin{\psi} \big) \hat{\mathbf{e}}_{\theta} + \big( h_\parallel \sin{\psi} + h_{\perp} \cos{\psi} \big) \hat{\mathbf{e}}_s\,,
\end{equation}
we obtain
\begin{equation}
    2 \eta u^{s \theta} = - \frac{\zeta \Delta \mu}{2 \sqrt{g}} \, \, \textrm{sin}\big(2 \psi \big) - \frac{\nu}{2 \sqrt{g}} h_\parallel  \textrm{sin}\big(2 \psi \big) - \frac{1}{2 \sqrt{g}} \Big( 1 + \nu \textrm{cos}\big(2 \psi \big) \Big) h_\perp 
\end{equation} By plugging in the value for $h_\parallel$ from \eqref{eq:strain_ustheta_relation}, we now obtain the expression for the strain rate in the form of $\psi$.
\begin{equation}
     u^{s\theta} = \frac{-1}{4 \eta \sqrt{g} \left( 1 + \frac{\nu^2 \gamma}{4 \eta}  \textrm{sin}^2{\big( 2\psi \big)} \right) }  \Big[  ( 1 + \nu \cos{2\psi} ) h_\perp + \zeta \Delta \mu \sin{2\psi}   \Big]. 
     \label{eq:ustheta}
\end{equation}
This allows us to eliminate $u^{s\theta}$ from \eqref{eq:simplified_rotational_dynamics}, and we thereby obtain following differential equation describing the dynamics of a +1 defect :
\begin{equation}
    \frac{\partial \psi }{\partial t}  = \Bigg[ \frac{1}{\gamma} +  \frac{(1+\nu \cos{2\psi})^2}{4 \eta  \left( 1 + \frac{\nu^2 \gamma}{4 \eta} \textrm{sin}^2{\big( 2\psi \big)}  \right) } \Bigg] h_\perp + \frac{1+\nu \cos{2\psi}}{4 \eta  \left( 1 + \frac{\nu^2 \gamma}{4 \eta} \textrm{sin}^2{\big( 2\psi \big)}  \right) }  \sin{ 2\psi }  \zeta  \Delta \mu . \label{eq:dynamics_of_a_defect}
\end{equation}
This equation can be written more compactly as
\begin{align}
\frac{\partial \psi }{\partial t} = \frac{1}{\tilde{\gamma}(\psi)} h_\perp+\frac{\zeta\Delta\mu}{4\tilde{\eta}(\psi)}(1+\nu\cos{2\psi})\sin{2\psi}\,,
\label{eq:PsiDynamics}
\end{align}
where 
\begin{equation}
\tilde{\gamma}(\psi)^{-1}= \frac{1}{\gamma} +  \frac{(1+\nu \cos{2\psi})^2}{4 \eta  \left( 1 + \frac{\nu^2 \gamma}{4 \eta} \textrm{sin}^2{\big( 2\psi \big)}  \right) }\hspace{1cm}{\rm and}
\hspace{30pt} \tilde{\eta}(\psi)=\eta  \left( 1 + \frac{\nu^2 \gamma}{4 \eta} \textrm{sin}^2{\big( 2\psi \big)}  \right)\,.
\end{equation}
Note that both $\tilde{\gamma}$ and $\tilde{\eta}$ are positive.
If we use the chosen form for the free energy $F$, which is defined in the main text as Eq.~2, the perpendicular field entering Eq.~\ref{eq:PsiDynamics} is given by 
\begin{equation}
    h_\perp = -\frac{\delta F}{\delta \psi}= K \nabla^2 \psi + K_{\rm{ex}} \, \Big( (C_\theta{}^\theta)^2 - (C_s{}^s)^2 \Big)  \frac{\textrm{sin}(2 \psi)}{2}  \,.
    \label{eq:h_perp}
\end{equation}
In Eq \eqref{eq:h_perp}, $\nabla^2 \psi $ is a Laplace-Beltrami operator, given by 
\begin{equation}
    \nabla^2 \psi = \frac{1}{r} \partial_s \Big( r \,   \partial_s \psi \Big)\,,
\end{equation}
which holds because of the assumed rotational symmetry of the problem.

\section{Determination of a velocity from $\psi$ field}
We next show how the velocity field $\mathbf{v}=v^\theta \mathbf{e}_\theta$ can be obtained from knowledge of the $\psi$.
First, assuming no-slip boundary condition at the surface edge, i.e., $v^\theta(s=s_{\rm max})=0$, integration of Eq.~\ref{eq:ShearStrainITOvtheta} gives
\begin{equation}
    v^\theta = \int_{s_{\rm max}}^s  2 u^{s\theta}\,ds'.
\end{equation}
Writing the flow field in terms of the physical velocity, that is, $\mathbf{v} = v(s,t) \, \hat{\mathbf{e}}_{\theta}$, we have
\begin{equation}
    v(s,t) = r(s) \int_{s_{\rm max}}^s ds' \, 2 u^{s\theta}.
\end{equation}
Next, to obtain the steady-state flow $v(s)$, we use Eqs.~\ref{eq:ustheta} and \ref{eq:PsiDynamics} to express
the flow in terms of $\psi$:
\begin{align}
		v(s) &= -\zeta\Delta\mu\,r\int_{s_{\rm max}}^s \frac{\tilde{\gamma}}{2\gamma\tilde{\eta}}\,\frac{\sin{2\psi}}{r^2} \sqrt{g} ds' \nonumber, \\
		&=-2\zeta\Delta\mu\,r\int_{s_{\rm max}}^s \frac{\sin{2\psi}}{r^2[\gamma\nu(\nu+2\cos{2\psi})+4\eta]} \sqrt{g} ds'\,.
\end{align}
This integral is evaluated numerically to determine the flow profiles; see Figs.~2, 3, 4 and S1, S3, S4, S5. 
\newpage
\section{Localized flows on a sphero-cylindrical surface}
\begin{figure}[h]
    \centering
    \includegraphics[width=
0.5\textwidth]{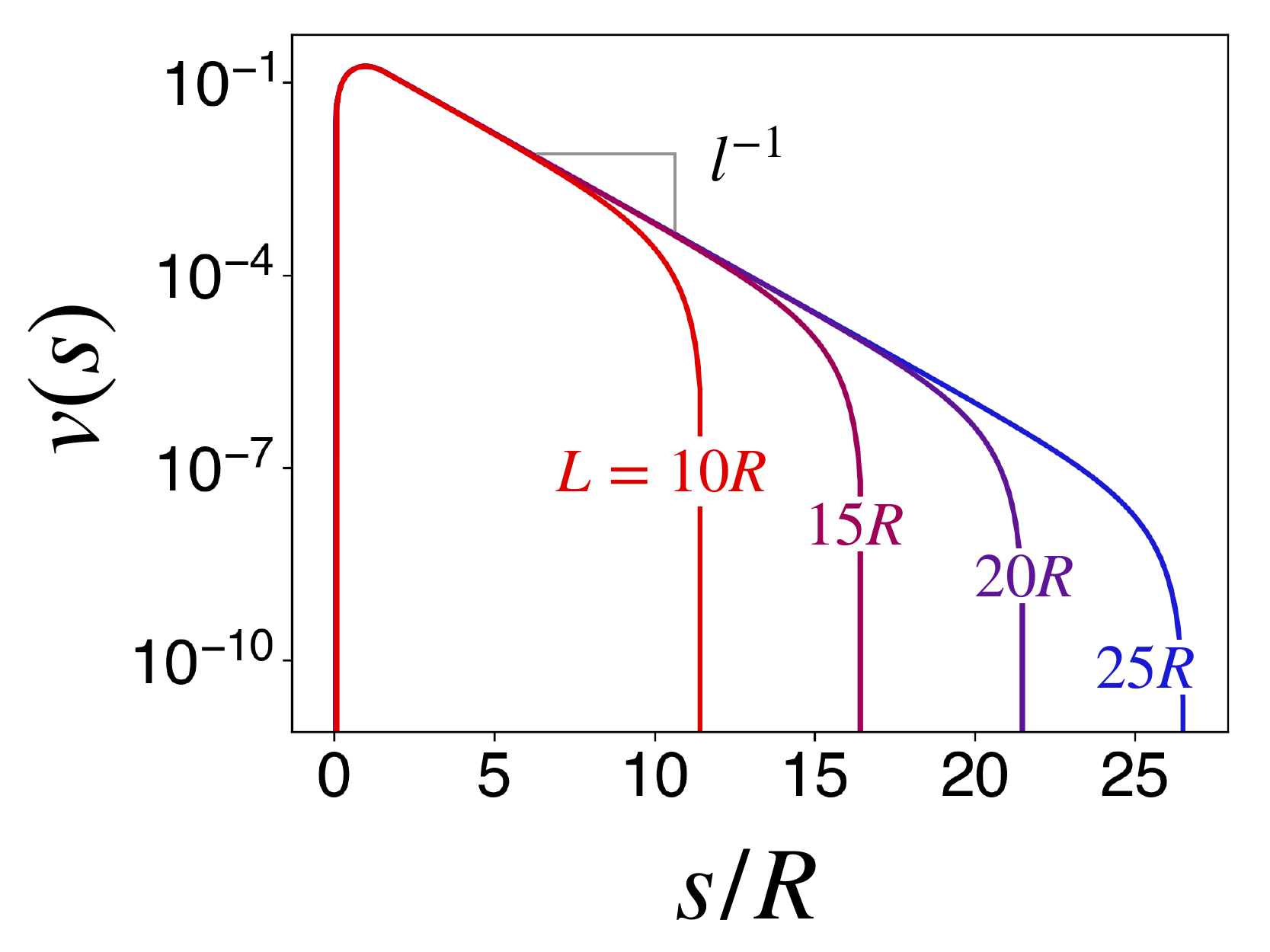}
    \caption{ \textbf{Localized flow on a sphero-cylindrical surface} : Flow profile, for extensile activity, for different cylinder lengths $L$, in units of the cylinder radius $R$. The velocity decays exponentially on the cylindrical surface. Parameters: $\mathcal{Z} = 1.1 \mathcal{Z}_c, R=1, K_{\rm{ex}}/K=1, \nu=-2, \eta=\gamma=1 $. Note: $\mathcal{Z}_c$ is defined in the main text.}
\label{fig:extensile_LC}
\end{figure}

Here we present numerical findings showing that the flows observed on a sphero-cylindrical surface decay exponentially. Moreover, if the length-scale of this decay (denoted as $l$) is small enough compared to the length of the cylinder (denoted as $L$),  we find that the flow profile is independent of $L$. In other words, the flow profile is insensitive to how far is the boundary ; see Fig.~ \ref{fig:extensile_LC}.

\section{Passive aster-spiral instability on a hemispherical bump surface}
We next present numerical results demonstrating that an aster can become passively unstable due to the influence of deviatoric curvature, $\mathcal{D}$, leading to a non-trivial spiral solution at equilibrium. In the absence of extrinsic coupling \emph{i.e.}, $K_{\rm{ex}}=0$, the steady-state passive $\psi$ dynamics equation $h_\perp =0$ simplifies to a diffusion problem. This admits a constant solution. The value of this solution is dictated by the boundary conditions.  However, the situation is strongly modified for $K_{\rm ex}\neq 0$.  

\begin{figure}[h!]
    \centering
    \includegraphics[width=
\textwidth]{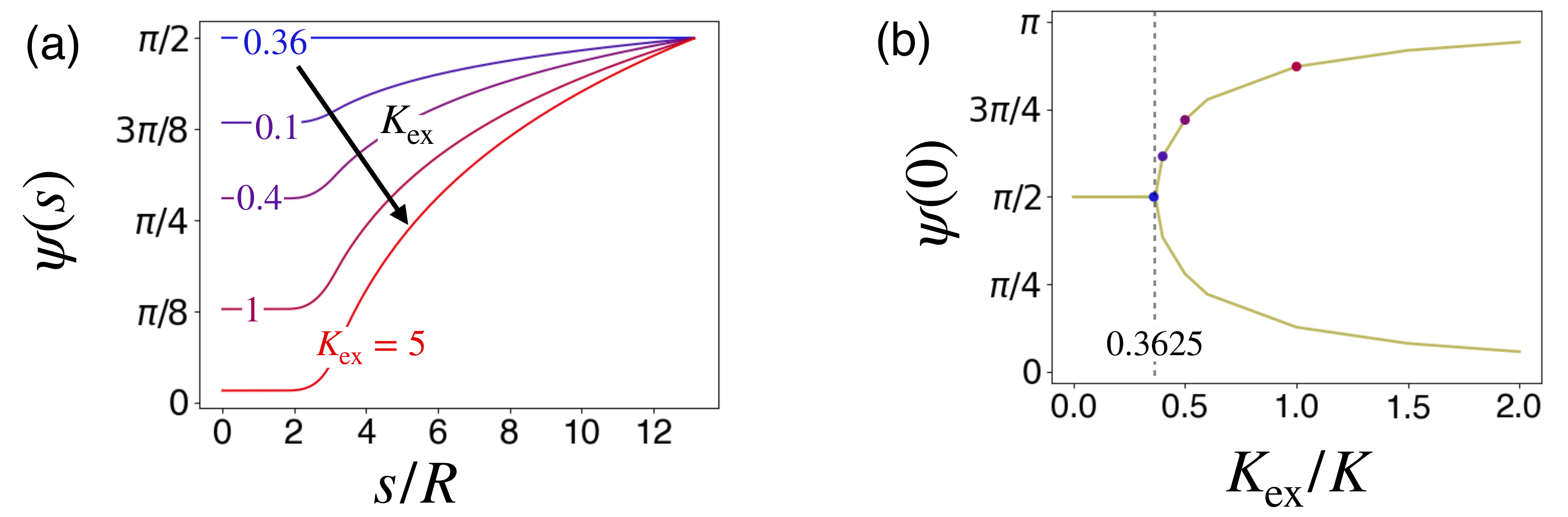}
    \caption{ \textbf{Supercritical aster-spiral transition at equilibrium.}  (a) $\psi$ profile for various values of $K_{\rm{ex}}$ on the hemispherical bump surface. Note: the threshold value of extrinsic coupling is $K_{\rm ex}=0.36$.  (b) Pitchfork bifurcation from aster to spiral for increasing $K_{\rm{ex}}$. Parameters : $\nu=-2, R=\epsilon=1, L=10R, K= 1 $.  }
\label{fig:passive_AS_transition}
\end{figure}

On the skirt region of the bump surface, $\mathcal{D}$ is negative. As discussed in the main text, for $K_{\rm{ex}}>0$, which is the case we consider throughout, the polarity $\mathbf{p}$ tends to align with $\mathbf{e}_\theta$. Therefore, assuming a boundary condition $\psi(s=s_{\rm max})=\pi/2$, the aster is stable. However, this constant solution can become unstable with sufficiently large extrinsic coupling.
In our numerics (solving Eq.~7 of the main text, with $\zeta\Delta\mu=0$), we observe a supercritical bifurcation from the aster to a spiral state (see Figs.~\ref{fig:passive_AS_transition}a and \ref{fig:passive_AS_transition}b). As the value of $K_{\rm{ex}}$ increases, the director field aligns more orthoradially. Consequently, for very high values of $K_{\rm{ex}}$ (or small $\epsilon$), the  orientational order resembles a vortex on the spherical dome.

\newpage

\section{Localized flows on the hemispherical bump surface }
\begin{figure}[h]
    \centering
    \includegraphics[width=
\textwidth]{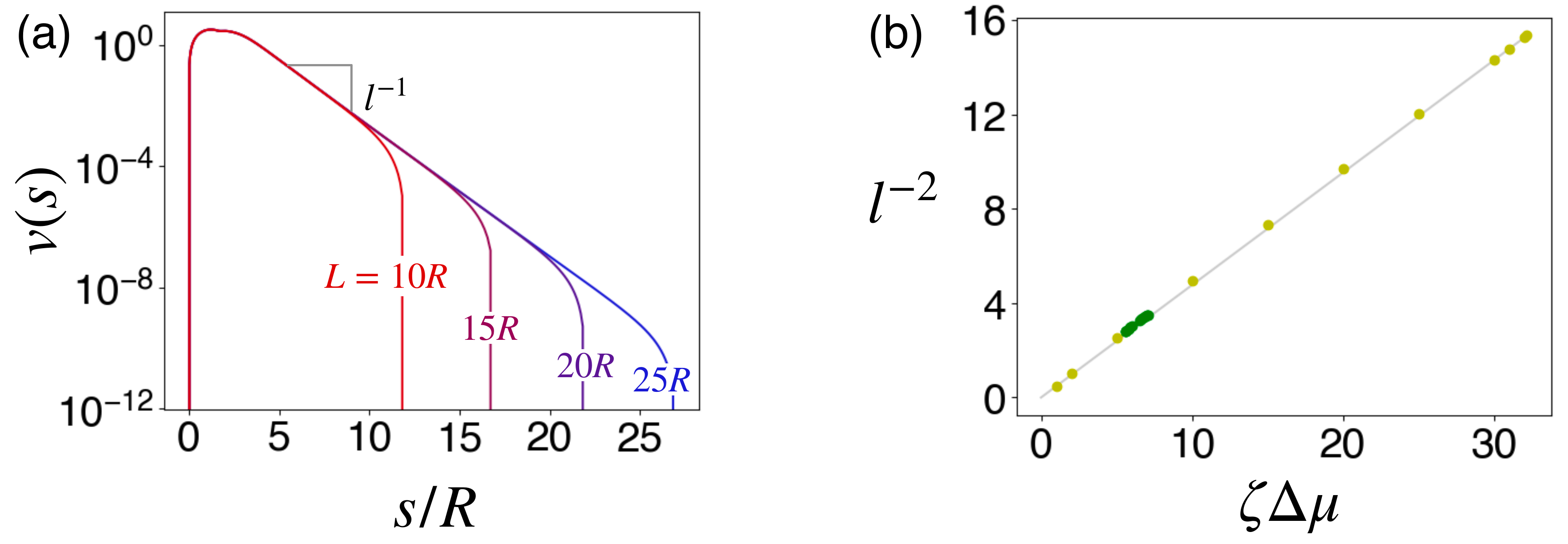}
    \caption{ \textbf{Contractility-driven localized flow on a hemispherical bump surface} : (a) Flow profiles for different values of  $L$. The velocity profiles decay exponentially on the flat surface. For $l\sim \sqrt{K/\zeta \Delta \mu} \ll L$,  $v(s)$ does not depend on $L$.  (b) The rate of the velocity decay is proportional to $\sqrt{\zeta \Delta \mu}$. The yellow points correspond to the yellow line (S1 solutions) and green points correspond to the green line (S2 solutions) in the hysteresis loop shown in Fig.~4a of the main text. }
\label{fig:contractile_LC}
\end{figure}

In this section we show that, in contrast to the case of a sphero-cylinder, active flows driven by contractility can also be localized : the velocity profile is insensitive to how far away the boundary is with respect to the bump skirt, and it exhibits an exponential decay. 
While bump surfaces with different values of $L$ (distance from outer skirt edge to outer boundary of flat surrounding region) will exhibit different passive spiral solutions, for high enough activity such that the length scale of velocity decay (denoted as $l$) is smaller than $L$, the localized flow profiles that are generated are  insensitive to the value of $L$; see Fig.~\ref{fig:contractile_LC}a. These flows become more pronounced and localized for increasing contractility, as exhibited by the scaling $l \propto 1/\sqrt{\zeta \Delta \mu}$; see Fig.~\ref{fig:contractile_LC}b.

\begin{figure}
    \centering
    \includegraphics[width=
0.4\textwidth]{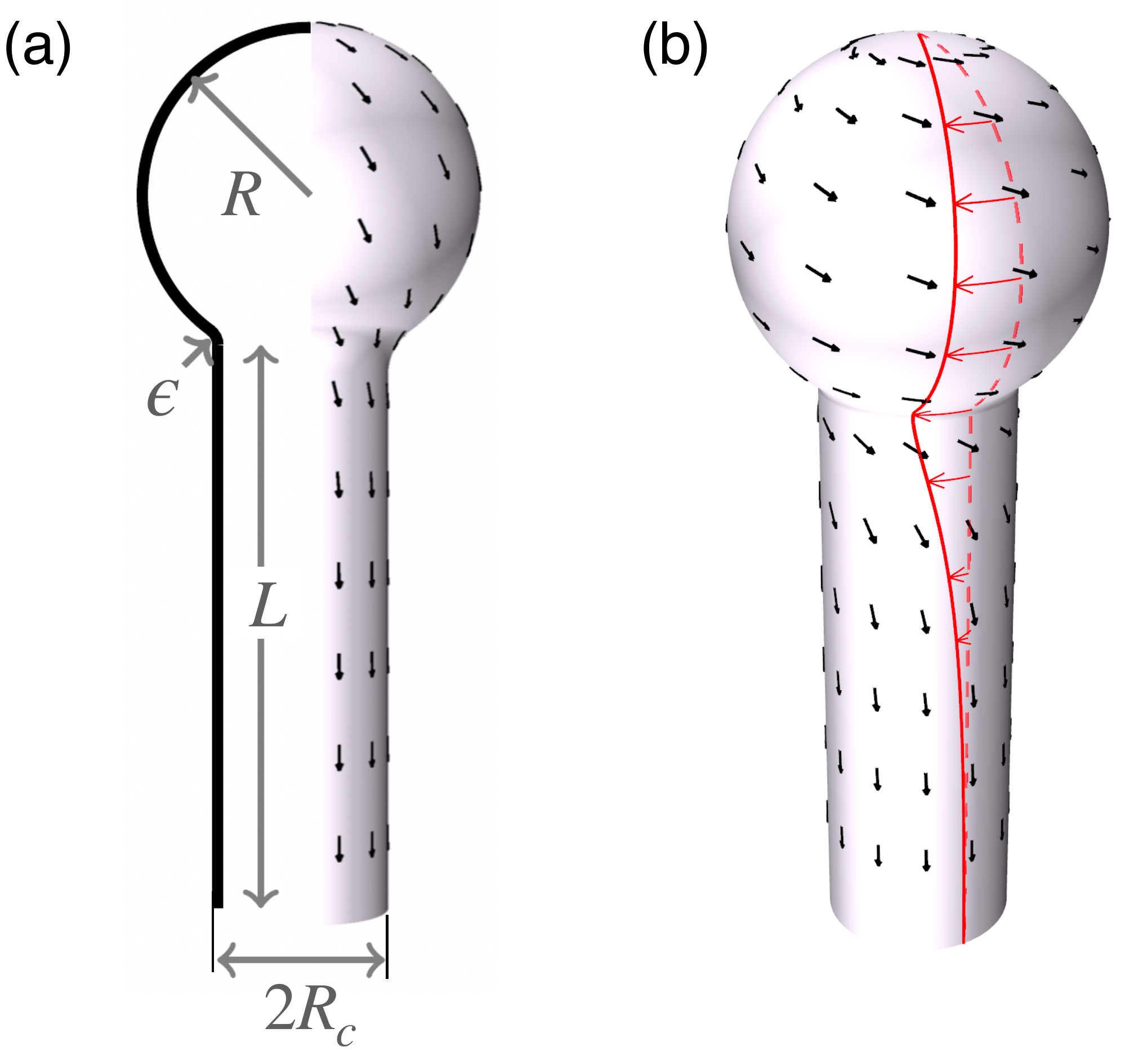}
    \caption{ \textbf{Localized flows for extensile activity on a bulb-tipped surface.} (a) Schematic representation of a bulb-tipped cylinder. This shows the non-uniform polarity profile (black arrows) at equilibrium. (b) Localized flow for extensile activity, $\zeta \Delta \mu<0$.  The velocity decays exponentially on the cylindrical surface. Parameters : $\zeta \Delta \mu=-2, R=1, \epsilon=0.1, R_c=0.546, L=7$. 
    }
\label{fig:extensile_LC_onpapillary}
\end{figure}

\newpage
\section{Results for other surfaces}


In this section, we present numerical results for surfaces other than those mentioned in the main text, and which illustrate that the findings obtained there are generic. 

\subsection{Bulb-tipped cylinder} We consider a bulbed-shaped surface, shown in Fig.~\ref{fig:extensile_LC_onpapillary}a, which is  combination of types (1) and (2) surfaces. It features a cylindrical base connected to a spherical top via a skirt.
Since $\mathcal{D}<0$ on the skirt, there is an energetic tendency for orthoradial alignment there.  Indeed, we observe a passive aster-to-spiral instability for a sufficiently small $\epsilon$, like we did for the hemispherical surface.  The non-uniform polarity in the passive state gives rise to threshold-less flows when activity is turned on. These are features we observed for the hemispherical bump (type 2 surface). 

Considering the case of $\zeta\Delta\mu<0$, we find that  flows driven by extensile activity on this surface are localized, 
in line with our findings on the sphero-cylindrical surface, that we identified as a type 1 surface;  see Fig.~\ref{fig:extensile_LC_onpapillary}b. This confirms that a cylinder-like surface, due to its positive deviatoric curvature, can screen active flows by forcing the orientation field to align along the cylinder's axis (for $K_{\rm ex}>0$). 
 
\begin{figure}[h]
    \centering
    \includegraphics[width=
\textwidth]{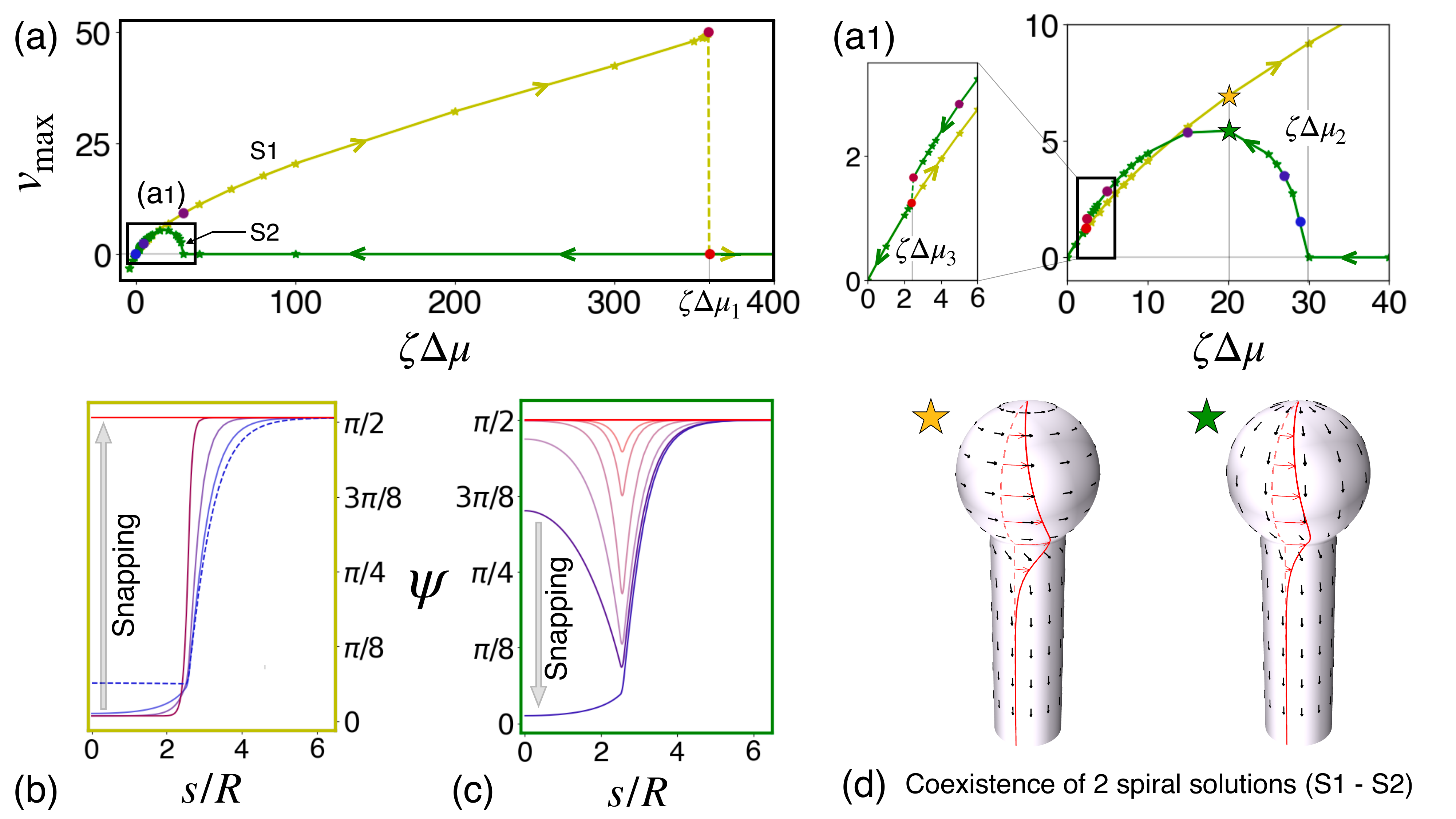}
    \caption{\textbf{Bistability of spiral and aster states on a bulb-tipped cylinder} (a) Hysteresis loop for $v_{\rm max}$ versus contractility. 
    (b) The $\psi(s)$ profiles for increasing contractility, color-coded according to the points on the yellow curve in (a), show a discontinuous spiral-to-aster transition.
    (c) The $\psi(s)$ profiles, color-coded according to the points on green curve (a1), show a continuous aster-to-spiral (S2) transition with decreasing $\zeta\Delta\mu$. Reducing the activity further causes a discontinuous spiral-to-spiral (S2 to S1) transition.
  (e)  Close-up of polarity and flow profiles for S1 and S2 spiral states. Parameter values :
    $R=1, \epsilon=0.1, R_c = 0.546, \nu=-2, K_{\rm{ex}}=K=1.$  
	}
    \label{fig:BulbTippedCylinder}
\end{figure}

 Next, considering the case $\zeta\Delta\mu>0$, we find that contractility also generates threshold-less, localized flows. High contractility extinguishes these flows and restores the aster, as we observed for the hemispherical bump; see Figs.~3, 4 of the main text and ~\ref{fig:BulbTippedCylinder}. For small $\epsilon$, the spiral-to-aster transition is discontinuous and we observed hysteresis in the $v_{\rm max}$ vs. $\zeta\Delta\mu$ plane, as we saw for the hemispherical bump; see Figs.~4 and~\ref{fig:BulbTippedCylinder}. 

\subsection{Bump-function} 
Finally, to illustrate the generality of the results we obtained in the main text for the fluid on a hemispherical bump, we consider a fluid on a surface of revolution described the so-called  \textit{bump function}, which is joined at its edge to a flat surface.  The particular bump function we chose is defined by the equation for the height at radial position $r$:
\begin{align}
z(r) &= H {\rm{exp}}\left( -\frac{1}{1-r^2} \right)    && 0 \leq r< 1 \\
z(r) &=0  && 1\leq r \leq 1+L\,,
\end{align}
where the height of the top of the bump is $z(0)=H/e$.  
We choose $H=e, L=3, K_{\rm ex}/K=1, K=1$. For this set of parameters, we readily observe passive aster-spiral instability. Similar to case of the hemispherical surface (main text), we observe localized flows for contractile activity, a discontinuous spiral-to-aster transition as well as bistability between two different flowing spiral solutions (S1 and S2) and a discontinuous transition from S2 to S1 (data not shown here).

